\documentstyle[12pt,chicagob]{article}

\advance \topmargin by -\headheight
\advance \topmargin by -\headsep

\evensidemargin \oddsidemargin

\newcommand{\blst}{\begin{list}{}{}}
\newcommand{\elst}{\end{list}}
\newcommand{\ben}{\begin{enumerate}}
\newcommand{\een}{\end{enumerate}}
\newcommand{\bit}{\begin{itemize}}
\newcommand{\eit}{\end{itemize}}
\newcommand{\bl}{\item}

\newcommand{\mfig}[1]{Figure~\ref{#1}}

\newcommand{\mthm}[1]{Theorem~\ref{#1}}

\newcommand{\msct}[1]{Section~\ref{#1}}

\newcommand{\vp}{\varphi}

\newcommand{\mr}[1]{\mbox{#1}}

\newcommand{\mb}[1]{{\bf #1}}

\newcommand{\emph}[1]{{\em #1\/}}

\newcommand{\Or}{\vee}

\newtheorem{THEOREM}{Theorem}[section]
\newenvironment{theorem}{\begin{THEOREM} \hspace{-.85em} {\bf :} }%
                        {\end{THEOREM}}
\newtheorem{LEMMA}[THEOREM]{Lemma}
\newenvironment{lemma}{\begin{LEMMA} \hspace{-.85em} {\bf :} }%
                      {\end{LEMMA}}
\newtheorem{COROLLARY}[THEOREM]{Corollary}
\newenvironment{corollary}{\begin{COROLLARY} \hspace{-.85em} {\bf :} }%
                          {\end{COROLLARY}}
\newtheorem{PROPOSITION}[THEOREM]{Proposition}
\newenvironment{proposition}{\begin{PROPOSITION} \hspace{-.85em} {\bf :} }%
                            {\end{PROPOSITION}}
\newtheorem{DEFINITION}[THEOREM]{Definition}
\newenvironment{definition}{\begin{DEFINITION} \hspace{-.85em} {\bf :} \rm}%
                            {\end{DEFINITION}}
\newtheorem{CLAIM}[THEOREM]{Claim}
\newenvironment{claim}{\begin{CLAIM} \hspace{-.85em} {\bf :} \rm}%
                            {\end{CLAIM}}
\newtheorem{EXAMPLE}[THEOREM]{Example}
\newenvironment{example}{\begin{EXAMPLE} \hspace{-.85em} {\bf :} \rm}%
                            {\end{EXAMPLE}}
\newtheorem{REMARK}[THEOREM]{Remark}
\newenvironment{remark}{\begin{REMARK} \hspace{-.85em} {\bf :} \rm}%
                            {\end{REMARK}}

\newcommand{\thm}{\begin{theorem}}
\newcommand{\lem}{\begin{lemma}}
\newcommand{\pro}{\begin{proposition}}
\newcommand{\dfn}{\begin{definition}}
\newcommand{\rem}{\begin{remark}}
\newcommand{\xam}{\begin{example}}
\newcommand{\cor}{\begin{corollary}}
\newcommand{\prf}{\noindent{\bf Proof:} }
\newcommand{\ethm}{\end{theorem}}
\newcommand{\elem}{\end{lemma}}
\newcommand{\epro}{\end{proposition}}
\newcommand{\edfn}{\bbox\end{definition}}
\newcommand{\erem}{\bbox\end{remark}}
\newcommand{\exam}{\bbox\end{example}}
\newcommand{\ecor}{\end{corollary}}
\newcommand{\eprf}{\bbox\vspace{0.1in}}
\newcommand{\beqn}{\begin{equation}}
\newcommand{\eeqn}{\end{equation}}

\newcommand{\bbox}{\vrule height7pt width4pt depth1pt}

\newcommand{\clm}{\begin{claim}}
\newcommand{\eclm}{\end{claim}}

\renewcommand{\phi}{\varphi}

\newcommand{\eg}{e.g.,~}
\newcommand{\ie}{i.e.,~}

\newcommand{\ol}{\setlength{\itemsep}{0pt}\begin{enumerate}}
\newcommand{\eol}{\end{enumerate}\setlength{\itemsep}{-\parsep}}
\newcommand{\ul}{\setlength{\itemsep}{0pt}\begin{itemize}}
\newcommand{\dl}{\setlength{\itemsep}{0pt}\begin{description}}
\newcommand{\edl}{\end{description}\setlength{\itemsep}{-\parsep}}
\newcommand{\eul}{\end{itemize}\setlength{\itemsep}{-\parsep}}

\newcommand{\commentout}[1]{}

\newcommand{\bi}{\begin{itemize}}
\newcommand{\ei}{\end{itemize}}
\newcommand{\be}{\begin{enumerate}}
\newcommand{\ee}{\end{enumerate}}

\newcommand{\js}{\vec{S}}

\def\bbbu{{\mathchoice {\setbox0=\hbox{$\displaystyle\rm U$}\hbox{\hbox
to0pt{\kern0.4\wd0\vrule height1\ht0\hss}\box0}}
{\setbox0=\hbox{$\textstyle\rm U$}\hbox{\hbox
to0pt{\kern0.4\wd0\vrule height1\ht0\hss}\box0}}
{\setbox0=\hbox{$\scriptstyle\rm U$}\hbox{\hbox
to0pt{\kern0.4\wd0\vrule height1\ht0\hss}\box0}}
{\setbox0=\hbox{$\scriptscriptstyle\rm U$}\hbox{\hbox
to0pt{\kern0.4\wd0\vrule height1\ht0\hss}\box0}}}}

\newcommand{\real}{\mb{R}}

\newcommand{\Tt}{\mb{true}}
\newcommand{\Ff}{\mb{false}}
\newcommand{\ep}{\mb{E}^\pay}
\newcommand{\pr}{\mr{Pr}}

\begin{document}

\title{On the NP-Completeness of 
Finding an Optimal Strategy in Games with Common Payoffs}
\author{Francis Chu%
\thanks{Work supported in part by NSF under
grant IRI-96-25901.}
\hspace{.5in}  Joseph Halpern%
\footnotemark[1]  \\ 
\\[.05in]
Department of Computer Science \\
Upson Hall, Cornell University \\
Ithaca, NY  14853-7501, USA \\
\\[.05in]
{\tt \{fcc,halpern\}@cs.cornell.edu}\\
}
\date{\today \\[.1in]
} %
\maketitle
\thispagestyle{empty}

\begin{abstract}
Consider a very simple class of (finite) games:
after an initial move by nature, each player makes one move.
Moreover, the players have common interests: at each node, all the
players get the same payoff.
We show that the problem of determining whether there
exists a joint strategy 
where each player has an expected payoff of at least $r$
is NP-complete
as a function of the number of nodes in the 
extensive-form
representation of the game.
\end{abstract}

\section{Introduction}

In many problems arising in distributed computing, we are free to
program the agents so as to achieve a goal of the designer; thus, the
agents can be viewed as pursuing a common goal.  
The interaction becomes a game in which the strategic aspect is of
little importance, whereas the coordination aspect becomes crucial.

To give just one example,
in recent years, the Federal Aviation Administration (FAA) has
introduced the 
concept of \emph{Free Flight}, which will decentralize
the National Airspace System (NAS).  
Free Flight 
allows pilots, whenever
practical, to choose their own route, instead of following pre-assigned routes.
(See {\tt http:/$\!$/www.faa.gov/freeflight/ff\_ov.htm} for more
 details.)
Since a pilot may plot his own course instead of taking a pre-assigned route
under Free Flight, the pilot might well want to optimize the route for his
payoff function.  
Of course, the optimal choices will depend both on what other pilots do
and on what can be viewed as Nature's moves (\eg the wind speed).
other pilots.  
as players in a game,  

While pilots are no longer obligated to follow routes pre-assigned by the FAA
during Free Flight, they might not want to act completely independently, since
there might be incentives to cooperate.  For example, pilots from the same
organization (\eg airline, shipping company, etc.) might well want to
coordinate their flights so as to optimize the payoff function of the
organization itself, since what is optimal for a particular flight might not be
the best thing to do for the organization as a whole.  
Indeed, in the extreme case, the pilots may have the same payoff
function (the payoff function of the organization to which they belong).

This type of of situation, where the agents have a common payoff because
they are  members of the same team or organization (or because their payoff can
be taken to be the organization's payoff) arises frequently
in distributed systems applications.
In games
of this sort, it turns out that there is an optimal 
joint
strategy---one that
gives 
all
the agents at least as high a payoff 
(individually)
as any other 
joint strategy.
This optimal
joint
strategy is deterministic and 
is Pareto optimal, a Nash equilibrium, and a correlated equilibrium
(under perhaps the  most natural definition of correlated equilibrium in
extensive-form games).

An obvious goal is thus to compute the optimal joint strategy.
Here, unfortunately, is where the 
scalability
concern bites.
We show that the problem 
of computing the optimal joint strategy
is NP-complete as a function of the number of nodes in the 
extensive-form 
representation of the game.  (More precisely, we show that it is 
NP-complete to determine whether there is a 
joint
strategy that nets the
players a payoff of 
at least $r$ for a fixed rational number $r$.)  
This is true even if there are only two players in the game, each
of whom moves once, after an initial move by nature.  
The role of nature
is critical.  Without the initial move by nature, it is easy to see
that the problem is decidable in 
polynomial time, 
no matter how many
players there are.  It is also easy to see that two players are required
for the lower bound; if there is just a single player, it is again easy
to figure out the optimal move for that player.

These results also depend on the fact that we are considering the
extensive-form representation of the game.  Using the 
normal-form 
representation
of the
game, it is trivial to find an optimal joint strategy for the players by
looking at the payoff matrix: it is just a joint strategy that nets them
the best payoff.  (Recall that we are considering games where the
players get the same payoff.)  There is no contradiction with the
NP-completeness result here: the normal-form representation can be 
exponential in the size of the extensive-form representation.

These results form an interesting contrast to those of Gilboa and Zemel
\citeyear{GZ89}, who considered related questions in the context of games
in 
normal form.  
They showed that, given a game $G$ and a number $r$,
computing whether there exists a Nash equilibrium where each player gets
a payoff of at least $r$ is NP-complete, while computing whether there
exists a correlated equilibrium where each player gets at least $r$ is
decidable in 
polynomial time.  
Note that Gilboa and Zemel are trying to
determine whether there exists a Nash or correlated equilibrium where
the players do well, whereas we are trying to determine whether there
exists a joint strategy where the players do well.  However, in our
setting, there exists a joint strategy where the players do well iff
there is a Nash/correlated equilibrium where the players do well.

It is also interesting to compare these results to those of Koller and
Megiddo \citeyear{KM92}.  Like us, they consider 
extensive form
and focus on the two-player case but they consider zero-sum
games, where the players have diametrically opposite interests, whereas
we consider coordination games where the players have identical
interests.  They show that it is NP-hard to decide whether player 1 
can {\em guarantee\/} an expected payoff of at least $r$ (no matter what
player 2 does), even if there are no chance moves.  They show the
problem is NP-complete if there are no chance moves or if player 1 has
perfect recall, but need to allow player 2 to have imperfect recall.
Note that since we restrict players to making one move each, they
certainly have perfect recall in our setting.

The rest of the paper is organized as follows.  
In \msct{sec:game} we 
briefly review the relevant definitions and explain the difficulty of
computing an optimal solution.
In \msct{sec:np}, we 
prove the main
NP-completeness result of the paper.  In \msct{sec:con} we offer some
concluding remarks.
\section{Preliminaries}
\label{sec:game}
\newcommand{\obs}{{\mbox{\sc obs}}}
\newcommand{\pay}{{\mbox{\sc pay}}}
\newcommand{\gm}{G}
\newcommand{\lpay}{\preceq}

We first describe the games of interest to us in somewhat nonstandard
notation (which nevertheless seems to us fairly natural and appropriate
for our application).  We then describe the easy conversions to normal
and 
extensive form.
For us, an $n$-player game, $\gm_n$, is a
$2n+3$-tuple $(W, A_1, \ldots, A_n, O_1, \ldots, O_n, \pr, \obs,
\pay)$ such that
\bit
\bl $W$ is the set of possible worlds,
\bl $A_i$ is the set of possible actions of agent $i$, 
\bl $O_i$ is the set of possible observations of agent $i$,
\bl $\pr$ is the probability distribution on $W$,
\bl $\obs(w, i) \in O_i$ is the observation agent $i$ makes in world $w$, and
\bl $\pay(w, (a_1, \ldots, a_n)) \in \real^n$ is the joint payoff of the joint action
$(a_1, \ldots, a_n)$ in world $w$.
\eit
Note that each world $w \in W$ determines the observation each agent makes (via
$\obs$) and the joint payoff given any joint action (via $\pay$).  We
assume 
that the game (\ie the tuple) is common knowledge among the agents.
Intuitively, if $\pay(w, (a_1, \ldots, a_n)) = (b_1, \ldots, b_n)$, then
$b_i$ is the payoff to agent $i$ if the world is $w$ and agent $j$
performs action $a_j$, $j=1, \ldots, n$. 
A {\em game with common payoffs\/} 
is one where all joint payoffs have
the form $(b, \ldots, b)$: \ie each agent gets the same payoff.

Each agent $i$ decides what action to take based on his observation and
his strategy.  
Formally, 
a \emph{strategy} for agent $i$ is a function from $O_i$ to $A_i$.
That is, for each observation, the strategy prescribes an action.  (Note that
this is a \emph{deterministic} strategy.  In general, \emph{randomized}
strategies are also possible.  However, 
as we show shortly,
in the games with common payoffs that we are interested in, there is no
loss of generality in considering
only on deterministic strategies.)
A \emph{joint strategy} is an
$n$-tuple $(S_1, \ldots, S_n)$ such that $S_i$ is the strategy for
agent $i$.  (For 
brevity, we will sometimes write $(x_1, \ldots, x_k)$ as $\vec{x}$.)  Given a
joint strategy $\js$, we can compute its expected joint payoff
\[
\ep(\js) = \sum_{w \in W} \pr(w) \cdot
\pay(w, (S_1(\obs(w, 1)), \ldots, S_n(\obs(w, n)))).
\]
\commentout{
Note that $S_i(\obs(w, i))$ is the action of agent $i$ in world $w$.  The above
is a slight abuse of notation, since random variables and their expected values
are typically real numbers.  What we really have is $n$ random variables and we
are computing each of their expected values, and we are expressing the
computation in terms of vector operations.  (Given the way scalar
multiplication and vector addition work, we get exactly what we want.)  If
the expected joint payoff of $\js = (S_1, \ldots, S_n)$ is $(p_1, \ldots,
p_n)$, we say that the expected payoff of agent $i$ is $p_i$ with respect to
$\js$, which we denote by $\ep_i(\js)$.
We can define a partial
ordering $\lpay$ on the joint strategies based on their expected joint payoffs
as follows:
$\js \lpay \js'$ iff $\ep_i(\js) \leq \ep_i(\js')$ for all $i$.
Note that if $\js \lpay \js'$, then $\js'$ is at least as good as $\js$ for all
the agents, since no one would be worse off by switching to $\js'$.
A joint strategy is \emph{Pareto optimal} if its expected joint payoff is
maximal with respect to $\lpay$.  
}%

Clearly such a game can be easily converted to 
extensive form.  
Nature
makes the first move, by choosing a world.  Then agent 1 moves (by
choosing an action in $A_1$), agent 2 moves (by choosing an action in
$A_2$), and so on.  Agent $i$ cannot distinguish two nodes $n$ and $n'$
in the game tree if nature's move in the path to $n$ is $w$, nature's
move in the path to $n'$ is $w'$, and $\obs(w,i) = \obs(w',i)$.  That
is, agent $i$'s information sets are determined by the $\obs(\cdot, i)$.
Note that the size of the game tree in the extensive-form representation
is essentially the same as the size of the description of the $\pay$
function.  Thus, the conversion from the representation that we have
chosen to the extensive-form representation is polynomial.

It is also easy to convert a game to 
normal form, 
by describing the
matrix of expected payoffs for each joint strategy.  However, in general, this
conversion is exponential.  For example, it is easy to construct a
two-player 
game where $|W| = n^2$, $|O_1|= |O_2| = n$, and $|A_1| = |A_2|
= n$.  The description of this game is quadratic in $n$, and the game tree
has $4n^2$ nodes.  However, since a strategy for player $i$ is a function
$O_i$ to $A_i$, there are $2^n$ strategies for each player, and the
normal-form representation of the game is of size exponential in $n$.

As is well known, finite games always have Pareto optimal joint
strategies and always have strategies that are in Nash equilibrium (if
we allow randomized strategies).  However, in general, the set of Pareto
optimal strategies may be disjoint from the strategies in Nash
equilibrium.  This is not the case with common payoffs.
Indeed, with common payoffs, 
$\lpay$ is a linear (pre)ordering, so there is
a joint strategy with the highest expected (joint) payoff.  (There may be more
than one, since there could be ties.)  Thus, there is always a
deterministic joint strategy
that dominates every joint strategy (with respect to the $\lpay$ ordering);
such a joint strategy is \emph{a fortiori} Pareto optimal and a Nash
equilibrium, since no agent can do (strictly) better with any other joint
strategy.  
Moreover, since the payoff of a randomized strategy is a
convex combination of the payoffs of deterministic strategies, there can
be no randomized strategy with a higher payoff.
Thus, in games with common payoffs, a (deterministic) joint strategy
with highest expected payoff 
is both a Nash equilibrium and Pareto optimal.

\commentout{
As suggested in the introduction, we 
can view Free Flight as a game, with the pilots as the players.
Each pilot acts according to his strategy and his observation.  We can
assume that the observation includes his position, velocity, wind speed,
and other such relevant factors (which, after all, the pilot learns by
observing his cockpit instruments).  Of course, the space of possible
observations is enormous, so computing the optimal strategy will be
nontrivial.  Indeed, as we show in the next section, it is NP-complete.
}
We are interested in characterizing the difficulty of computing the
optimal strategy.  To make this a decision problem, we consider the
problem of deciding whether the optimal strategy gives a payoff of at
least $r$.  As we said in the introduction, we show that this problem is
NP-complete for our representation (and hence the extensive-form
representation) of the game.  Why should it be so hard?

Clearly, it is trivial in the case of normal-form games: we simply scan
the possible payoffs and find whether there is one that gives an
expected payoff of at least $r$.  However, since the normal-form
representation is 
exponential in the size of the extensive-form representation, this will
not help.   If there were no chance moves, the problem would also be
trivial.  We simply scan the payoffs; if there is one with a payoff of
at least $r$, the answer is yes, since the players can just play the
strategy that puts them on this path.  However, with chance moves, this
approach will not work, since we must take nature's move into account.  
This turns out to be hard, even if nature has only polynomially many
possible moves. 

\section{The NP-Completeness Result}
\label{sec:np}

We now show the problem of determining the optimal strategy in a game
with common payoffs is NP-complete.  For simplicity, we restrict to
two-player
games.  It will be clear from the proof that the result holds
for arbitrary $n$-player games.
For technical reasons, we 
further restrict to games
whose probabilities and payoffs are rational numbers, and that the rational
number $\frac{p}{q}$ is represented by pairs of integers $(p,q)$, where $q >
0$.  (This ensures that the games are finitely \emph{described}, so that we can
finish reading their descriptions in finite amount of time.  If we are truly
pedantic, we should actually fix an encoding of the games, but we will not get
into such tedious details.  Readers not familiar with NP-completeness results
and the techniques by which they are established may wish to consult
\cite{GarJoh} for an introduction.)
If the game is a common payoff game, we 
write $\pay(w, (a_1, \ldots, a_n)) = p$, where $p$ is the common payoff.
We then take $\ep(\js)$ to be a single number, rather than a tuple (all
of whose components are identical).

The problem of finding the optimal strategy is an optimization problem.
So that it fits into the standard framework of complexity classes, we
convert it to a decision problem:
\begin{quote}
\begin{list}{}{
\setlength{\labelwidth}{2in}
}
\item[UTIL:] Given a 
two-player 
game (with common payoff), is there a joint strategy
$\js$ such that $\ep(\js) \geq 1$?%
\footnote{Of course, there is nothing special about the choice of 1
here.  The same result holds for an arbitrary 
(rational)
$r$ or if we take $r$ as
part of the input.
In general, for hardness results, it suffices to focus on a special case.}
\end{list}
\end{quote}
The following theorem shows that UTIL is NP-complete.

\thm
\label{thm:np}
UTIL is NP-complete.
\ethm
\prf
UTIL is clearly in NP, since it suffices to guess a joint strategy and verify
that the expected joint payoff is indeed at least $1$.  To show that UTIL is
NP-hard, we reduce 3SAT to UTIL.  Recall that an \emph{instance} of 3SAT
consists of a Boolean formula which is a conjunction of \emph{clauses}, each of
which is a disjunction of three \emph{literals}, each of which is either a
(Boolean) variable or the negation of a (Boolean) variable.  An instance of
UTIL is simply a (finite) 
two-player 
game (with common payoff).  A
\emph{positive instance of 3SAT} is a formula (of 3SAT) which is satisfiable
(\ie there is a truth assignment that satisfies all the clauses) while a
\emph{positive instance of UTIL} is a (finite) 
two-player 
game (with common
payoff) for which there exists a joint strategy whose expected payoff is at
least $1$.  Recall that to show that 3SAT reduces to UTIL, we need to give a
polynomial-time 
transformation $f$ that maps instances of 3SAT to
instances of UTIL such that $\vp$ is a positive instance of 3SAT iff
$f(\vp)$ is a positive instance of UTIL.

Let $\vp$ be an instance of 3SAT with $n$ clauses.
Let $n$ be the number of clauses in $\vp$ and
let $m$ be the number of variables in $\vp$.  (Note that $m \leq 3n$, since each
clause contains at most three variables.)  Let $z_1, \ldots, z_m$ be the
variables and $C_1, \ldots, C_n$ be the clauses.  Thus 
$\vp = C_1 \wedge \cdots \wedge C_n$, where
$C_i = (\ell_{i,1} \vee \ell_{i,2} \vee \ell_{i,3})$ is a 
clause, where $\ell_{i,j}$ is a literal.  
If $\ell_{i,j} = z_k$ or $\ell_{i,j} = \neg{z}_k$, we say that $z_k$ is the
variable 
\emph{associated with} $\ell_{i,j}$; we denote the variable associated with a
literal $\ell$ by $v(\ell)$. 
Basically, the game proceeds as follows: nature chooses a clause $C_i$ and
literal $\ell_{i,j}$ in that clause.  Each of the $3n$ choices of nature
is equally likely.  Player 1 observes the variable
$v(\ell_{i,j})$ associated with the literal chosen by nature, and  must
then choose a truth value for that variable.  Note that a strategy for
player 1 is a truth assignment.  Optimal play by player 1
will amount to choosing a truth assignment that satisfies $\vp$.
Player 2 observes the clause $C_j$ chosen by nature and must choose a
literal  in that clause. Intuitively, this should be a literal
that evaluates to 
$\Tt$
in the truth assignment chosen by player 1.
The players get a payoff of 3 if the literal chosen
by player 2 is the same as the one chosen by nature and it
evaluates to 
$\Tt$
under the truth value for the
variable chosen by player 1; otherwise they get 0.  

Notice that the maximum expected payoff the players can get in this game
is 1, since for each clause, in exactly one of the three worlds
corresponding to that clause, player 2's choice will match nature's
choice.  In the $n$ worlds where player 2's choice and nature's choice
match, the players can get a payoff of 3 if the player 2's choice
evaluates to true under player 1's truth assignment.  In all other
worlds, they get a payoff of 0.  Moreover, if $\vp$ is satisfiable, 
there is a joint strategy that gives the players an
expected payoff of 1.  Player 1 chooses a truth assignment
$\alpha$ that satisfies $\vp$ (so that when player 1 observes the
variable $z$, he plays $\alpha(z)$) and player 2, when given clause
$C_i$, chooses a literal $j$ in $C_i$ that evaluates to true under truth
assignment $\alpha$.  On the other hand, if $\vp$ is not satisfiable,
there is no joint strategy that nets an expected payoff of 1, since
for each truth assignment chosen by player 1, there is at least one
clause where no literals are satisfied.

More formally, 
let $f(\vp)$ be the following game:
\bit
\bl $W = \{w_{i,j} : 1 \leq i \leq n$ and $1 \leq j \leq 3\}$ 
(the world $w_{i,j}$ corresponds to the literal $\ell_{i,j}$),
\bl $A_1 = \{\Tt, \Ff\}$, 
\bl $A_2 = \{1, 2, 3\}$, 
\bl $O_1 = \{z_1, \ldots, z_m\}$,
\bl $O_2 = \{C_1, \ldots, C_n\}$, and
\bl $\pr(w_{i,j}) = \frac{1}{3n}$ for all $i,j$. 
\bl $\obs(w_{i,j}, 1) = v(\ell_{i,j})$ and $\obs(w_{i,j}, 2) = C_i$; that is,
in world $w_{i,j}$, the observation of agent 1 is $v(\ell_{i,j})$ and the
observation of agent 2 is $C_i$.  
\bl $\pay$ is defined as follows:
\[
\begin{array}{lcl}
\pay(w_{i,j}, (\Tt, j')) & = & \left\{
\begin{array}{ll}
3 & \mbox{if $j = j'$ and $\ell_{i,j} = v(\ell_{i,j})$} \\ 
0 & \mbox{otherwise} \\ 
\end{array}
\right.\\
\\
\pay(w_{i,j}, (\Ff, j')) & = & \left\{
\begin{array}{ll}
3 & \mbox{%
if 
$j = j'$ and 
$\ell_{i,j} = \neg v(\ell_{i,j})$} \\ 
0 & \mbox{otherwise} \\ 
\end{array} 
\right.\\
\end{array}
\]
\eit
Note that the size of $f(\vp)$ is linear in the number of clauses, so it is
easy to implement $f$ in 
polynomial time.  
As an example,
let $\vp = (z_1 \Or \neg{z}_2 \Or z_3) \wedge (z_2 \Or z_4 \Or
\neg{z}_1)$.  Then $f(\vp) = (W, A_1, A_2, O_1, O_2, \pr, \obs, \pay)$, where
\bit
\bl $W = \{ w_{1,1}, w_{1,2}, w_{1,3}, w_{2,1}, w_{2,2},  w_{2,3}\}$, %
\bl $A_1 = \{\Tt, \Ff\}$, %
\bl $A_2 = \{1, 2, 3\}$, %
\bl $O_1 = \{z_1, z_2, z_3, z_4\}$, %
\bl $O_2 = \{(z_1 \Or \neg{z}_2 \Or z_3),  (z_2 \Or z_4 \Or \neg{z}_1)\}$, and %
\bl $\pr(w_{i,j}) = \frac{1}{6}$. %
\bl $\obs$ 
is defined as follows:
\bit
\bl
$\obs(w_{1,1}, 1) = z_1$, 
$\obs(w_{1,2}, 1) = z_2$, 
$\obs(w_{1,3}, 1) = z_3$, \\
$\obs(w_{2,1}, 1) = z_2$, 
$\obs(w_{2,2}, 1) = z_4$, 
$\obs(w_{2,3}, 1) = z_1$
\bl
$\obs(w_{1,*}, 2) = (z_1 \Or \neg{z}_2 \Or z_3)$, \\
$\obs(w_{2,*}, 2) = (z_2 \Or z_4 \Or \neg{z}_1)$
\eit
\bl $\pay$
is defined as follows:
\begin{center}
\begin{minipage}{4in}
\begin{tabular}{c|c|c|c|}
 \multicolumn{1}{c}{} & \multicolumn{1}{c}{1} & \multicolumn{1}{c}{2} &
\multicolumn{1}{c}{3} \\
\cline{2-4}
 $\Tt$ & 3 & 0 & 0 \\
\cline{2-4}
 $\Ff$ & 0 & 0 & 0 \\
\cline{2-4}
\multicolumn{1}{c}{} & \multicolumn{3}{c}{$w_{1,1}$}\\
\end{tabular}
\hfill
\begin{tabular}{c|c|c|c|}
 \multicolumn{1}{c}{} & \multicolumn{1}{c}{1} & \multicolumn{1}{c}{2} &
\multicolumn{1}{c}{3} \\
\cline{2-4}
 $\Tt$ & 0 & 0 & 0 \\
\cline{2-4}
 $\Ff$ & 0 & 3 & 0 \\
\cline{2-4}
\multicolumn{1}{c}{} & \multicolumn{3}{c}{$w_{1,2}$}\\
\end{tabular}
\hfill
\begin{tabular}{c|c|c|c|}
 \multicolumn{1}{c}{} & \multicolumn{1}{c}{1} & \multicolumn{1}{c}{2} &
\multicolumn{1}{c}{3} \\
\cline{2-4}
 $\Tt$ & 0 & 0 & 3 \\
\cline{2-4}
 $\Ff$ & 0 & 0 & 0 \\
\cline{2-4}
\multicolumn{1}{c}{} & \multicolumn{3}{c}{$w_{1,3}$}\\
\end{tabular}
\end{minipage}
\end{center}
\begin{center}
\begin{minipage}{4in}
\begin{tabular}{c|c|c|c|}
 \multicolumn{1}{c}{} & \multicolumn{1}{c}{1} & \multicolumn{1}{c}{2} &
\multicolumn{1}{c}{3} \\
\cline{2-4}
 $\Tt$ & 3 & 0 & 0 \\
\cline{2-4}
 $\Ff$ & 0 & 0 & 0 \\
\cline{2-4}
\multicolumn{1}{c}{} & \multicolumn{3}{c}{$w_{2,1}$}\\
\end{tabular}
\hfill
\begin{tabular}{c|c|c|c|}
 \multicolumn{1}{c}{} & \multicolumn{1}{c}{1} & \multicolumn{1}{c}{2} &
\multicolumn{1}{c}{3} \\
\cline{2-4}
 $\Tt$ & 0 & 3 & 0 \\
\cline{2-4}
 $\Ff$ & 0 & 0 & 0 \\
\cline{2-4}
\multicolumn{1}{c}{} & \multicolumn{3}{c}{$w_{2,2}$}\\
\end{tabular}
\hfill
\begin{tabular}{c|c|c|c|}
 \multicolumn{1}{c}{} & \multicolumn{1}{c}{1} & \multicolumn{1}{c}{2} &
\multicolumn{1}{c}{3} \\
\cline{2-4}
 $\Tt$ & 0 & 0 & 0 \\
\cline{2-4}
 $\Ff$ & 0 & 0 & 3 \\
\cline{2-4}
\multicolumn{1}{c}{} & \multicolumn{3}{c}{$w_{2,3}$}\\
\end{tabular}
\end{minipage}
\end{center}
\eit
\commentout{
The intuition behind $f$ is easy to explain.  Since we want to map positive
instances of 3SAT to positive instances of UTIL, we want to establish a
correspondence between satisfying truth assignments and joint strategies whose
expected payoff is at least $1$, so that there exists a satisfying assignment
iff there exists a joint strategy with the desired expected payoff. 
The observations of agent 1 are variables and the actions of agent 1 are
truth values; thus, the strategies of agent 1 are truth assignments.
Intuitively, we can think of agent 1 as a ``guesser'' (who guesses a truth
assignment) and agent 2 as a ``verifier'' (who tries to verify that the guess
indeed satisfies the formula).  To try to verify that a guess satisfies the
formula, agent 2 picks a literal from each clause and sees if that literal is
satisfied by the truth assignment.  If so, the (joint) payoff in the particular
world corresponding to the literal is $3$ (recall that the world $w_{i,j}$
corresponds to the literal $\ell_{i,j}$); otherwise, the joint payoff is $0$.
It is easy to check that if every literal agent 2 picks is indeed satisfied by
the truth assignment, then the expected payoff is exactly $1$; otherwise the
expected payoff is less than $1$.
}

As discussed earlier,
it is easy to see that there is a joint strategy that gives the players
an expected payoff of 1 in this game iff $\vp$ is satisfiable.
Thus $\vp$ is a positive instance of 3SAT iff $f(\vp)$ is a positive
instance of UTIL, so we are done. \eprf
Although we restricted to 
two-player
games in Theorem~\ref{thm:np},
it should be clear that the analogous problem is also NP-complete for
$n$-player
games.  The upper bound is clear, since it suffices to guess a
joint strategy just as before.  And it is easy to modify our lower bound
proof to deal with $n$-player games; we leave details to the reader.

There is one technical point worth observing.
Say that player $i$ \emph{considers $w_j$ possible in $w_k$} iff
$\obs(w_k, i) = \obs(w_j, i)$ (\ie player $i$ makes the same observation in
both worlds).  Intuitively, if a player considers many worlds possible, he has
a lot of uncertainty.  Note that in the game constructed in the proof of
\mthm{thm:np}, player 2 considers only three worlds possible in any given world
(since there are only three literals in each clause) while player 1 may
consider many worlds possible in some worlds (since a variable may appear in
many clauses).  Is it necessary for one of the players to have much uncertainty
for our result to hold?  It turns out that the problem remains NP-complete even
if player 1 considers at most three worlds possible in each world as well.  The
reason that player 1 may consider many worlds possible is because a variable
may occur in many clauses.  It is easy to convert a formula $\vp$ in which a
variable may occur many times to a formula ${\vp'}$ in which each variable
occurs in at most three clauses such that ${\vp'}$ is satisfiable iff $\vp$ is
satisfiable and ${\vp'}$ can be constructed in 
polynomial time. 
(This can be
done via a technique very similar to the reduction of SAT to 3SAT.)  Thus the
problem remains NP-complete even if both players consider at most three worlds
possible in each world.  While we know that 2SAT is in P and that SAT
restricted to formulas in which each variable occurs at most twice is in P, we
have not investigated whether UTIL remains NP-complete if both players consider
at most two worlds possible in each world.  Clearly if both players consider
only one world possible (\ie they have perfect information), then we can find
the optimal joint strategy in linear time.

\section{Conclusion}
\label{sec:con}

We have shown that the problem of determining whether there is a joint
strategy that nets at least $r$ in a 
common payoff game in 
extensive form is
NP complete,
even if the there are only two players, each of whom makes only one move
(following a move by nature).  
Essentially the same argument
shows that it is NP-complete to find a strategy that is within a fixed
fraction of optimal.  Thus, we cannot even find approximately optimal
strategies in 
polynomial time.
What does this say about problems such as Free Flight?
Should we necessarily give up on finding optimal strategies?
Recent successes in finding solutions to NP-complete problems
\cite{Hogg96,Monasson99} suggest that there may be some reason to hope;
NP-complete problems may not be so infeasible in practice.  Of course,
further research needs to be done to see if problems such as Free Flight
can in fact be represented in a reasonable way as a game that is in
practice soluble.  

\section*{Acknowledgements}

We would like to thank Ken Birman for bringing Free Flight to our
attention, 
and an associate editor of the journal for a very thoughtful
review, and many useful suggestions for improving the paper.

\bibliographystyle{chicago}
\bibliography{z,joe}

\begin{thebibliography}{}

\bibitem[\protect\citeauthoryear{Garey and Johnson}{Garey and
  Johnson}{1979}]{GarJoh}
Garey, M. and D.~S. Johnson (1979).
\newblock {\em Computers and Intractability: A Guide to the Theory of
  {NP}-completeness}.
\newblock San Francisco, Calif.: W. Freeman and Co.

\bibitem[\protect\citeauthoryear{Gilboa and Zemel}{Gilboa and
  Zemel}{1989}]{GZ89}
Gilboa, I. and E.~Zemel (1989).
\newblock Nash and correlated equilibrium: some complexity considerations.
\newblock {\em Games and Economic Behavior\/}~{\em 1}, 80--93.

\bibitem[\protect\citeauthoryear{Hogg, Huberman, and Williams}{Hogg
  et~al.}{1996}]{Hogg96}
Hogg, T., B.~Huberman, and C.~Williams (Eds.) (1996).
\newblock {\em Artificial Intelligence}, Volume~81.
\newblock Elsevier.
\newblock Special Issue on {\em Phase Transitions and Complexity}.

\bibitem[\protect\citeauthoryear{Koller and Megiddo}{Koller and
  Megiddo}{1992}]{KM92}
Koller, D. and N.~Megiddo (1992).
\newblock The complexity of two-person zero-sum games in extensive form.
\newblock {\em Games and Economic Behavior\/}~{\em 4\/}(4), 528--552.

\bibitem[\protect\citeauthoryear{Monasson, Zecchina, Kirkpatrick, Selman, and
  Troyansky}{Monasson et~al.}{1999}]{Monasson99}
Monasson, R., R.~Zecchina, S.~Kirkpatrick, B.~Selman, and L.~Troyansky (1999).
\newblock Typical-case complexity results from a new type of phase transition.
\newblock {\em Nature\/}~{\em 400\/}(8), 133--137.

\end{thebibliography}

\end{document}